\documentclass[10pt]{article}
\usepackage{graphics}
\usepackage{ulem}
\usepackage{bm}
\newcommand{\bea}{\begin{eqnarray}}
\newcommand{\eea}{\end{eqnarray}}
\newcommand{\be}{\begin{equation}}
\newcommand{\ee}{\end{equation}}
\newcommand{\f}[1]{\frac{1}{2}}
\def\be{\begin{eqnarray}}
\def\ee{\end{eqnarray}}
\def\bd{\begin{displaymath}}
\def\ed{\end{displaymath}}
\def\nn{\nonumber}

\def\etal{{\it et al.}}

\def\PR{Phys. Rev. }
\def\PRL{Phys. Rev. Lett. }
\def\PL{Phys. Lett. }

\begin{document}
\centerline{\large Ground states and excited states of hypernuclei } 
\centerline{\large in Relativistic Mean Field approach}
\vskip 0.3cm
\centerline{ Bipasha Bhowmick, Abhijit Bhattacharyya and G. Gangopadhyay}
\vskip 0.3cm
\centerline{Department of Physics, University of Calcutta}
\centerline{92, Acharya Prafulla Chandra Road, Kolkata-700 009, India}
\centerline{email: ggphy@caluniv.ac.in}
\vskip 0.3cm

\begin{abstract}
 Hypernuclei have been studied  within the framework of Relativistic Mean 
Field theory. The force FSU Gold has been extended to include hyperons. 
The effective hyperon-nucleon and nucleon-nucleon interactions have
been obtained by fitting experimental energies in a number of hypernuclei 
over a wide range of mass. 
Calculations  successfully describe various features including hyperon
separation energy and single particle spectra of single-$\Lambda$ hypernuclei
 throughout the periodic table. 
We also extend this formalism to double-$\Lambda$ hypernuclei.
\end{abstract}
\section{Introduction}
\label{intro}

Hypernuclei are  the first kind of flavoured nuclei in the direction of other exotic systems.
 One of the main reasons of interest in hypernuclear physics lies in the characteristics
 of the hyperon-nucleon(YN) and hyperon-hyperon(YY) interactions, 
 crucial inputs to describe the structure of these strange nuclei.
Obviously, measurements 
 of YN and YY cross sections would give direct information on the interactions. However,
 such experiments are very difficult due to the short lifetime of the hyperons;
 till date no scattering data are available
 on the YY interaction while very limited data are available for the $\Xi$N
 interactions. 

As the $\Lambda$ is the lightest among the hyperons,  $\Lambda$ hypernuclei have been 
investigated more thoroughly than similar systems. Extensive experimental studies involving ($\pi^{+}$,$K^{+}$), 
(e,$e^{\prime}K^{+}$),  ($K^{-}$,$\pi^{-}$) or ($\gamma$,$K^{+}$) reactions have measured the 
binding energy, shell structure and other properties of single-$\Lambda$ hypernuclei over a 
wide range of the periodic table, (see, e.g. Hashimoto \etal\cite{1} for a recent review of the 
experimental scenario). Considerable amount of details about the $\Lambda$N interaction have already been extracted. 
 For example, It has been established\cite{2a,2b} 
that the spin-orbit part of
this force is weaker 
than that of the NN system. 

On the other hand, existing experimental information about $\Xi$ 
or  double-$\Lambda$ hypernuclear systems are extremely insufficient to draw any 
strong conclusion about $\Xi$N or the $\Lambda\Lambda$ interactions, respectively. 
Though a large amount of theoretical studies 
have been performed already to describe these systems (see,  for example, some Relativistic
Mean Field (RMF) works\cite{3,4,5})
the situation is not clearly
understood, especially due to the lack of experimental data. In this regard
the importance of these systems is increasing 
even more now-a-days with the program for search of H-dibaryon at J-PARC and also the 
SKS and the proposed S-2S experimental facilities at KEK/J-PARC concentrating especially 
on $\Xi$ hypernuclei.\cite{6}

It is, therefore, important to develop reliable theoretical tools to investigate the structure of
these systems. Both relativistic and non-relativistic descriptions were used for the purpose. 
SU$(3)$ symmetric field theories\cite{7a,7b} and the 
quark-meson coupling model\cite{8a,8b,8c}
were developed to investigate hypernuclei. A density dependent relativistic
hadron (DDRH) field theory was used by Keil \etal\cite{9} to describe $\Lambda$ hypernuclei. 
Among the non-relativistic approaches, one can name 
the shell model calculation,\cite{10} the semi-empirical mass formulas,\cite{11a,11b,11c} 
the phenomenological single-particle fields,\cite{12a,12b}
and the Skyrme-Hartree-Fock model (SHF),\cite{13,14} etc.

The RMF approach, which is very useful in describing the properties of normal nuclei,
 was used to study hypernuclear systems in various works.\cite{3,4,5,15a,15b,15c}
 The $\Lambda$N interaction can either be extracted 
from microscopic methods like G-matrix calculations, or adjusted by fitting the experimental data.
 In this work, we perform a study of $\Lambda$ hypernuclei within the framework of the RMF theory
using the FSU Gold Lagrangian density\cite{17} where the parameters of the hyperon
interaction 
are determined by fitting the experimental separation energies of several
hypernuclei in the mass 
region $16$ to $208$. 
Ideally, one should have looked to minimize the $\chi^2$ value. 
However, for a number of nuclei, the theoretical binding energy values from 
mean field calculations are not expected to be sufficiently accurate to 
approach the experimental values within the experimental error. Thus,
a $\chi^2$ minimization will lead to over-dependence on only a few
experimental values, particularly in the lighter hypernuclei.   
Hence, the parameter set that produces the minimum root mean square
(rms) deviation for the experimental separation
energies are adopted to calculate various other properties of hypernuclei
throughout the periodic table, and compared with experimental data, whenever 
possible. Unless otherwise mentioned, all our results are obtained from 
this parameter set. 

We extend this formalism to the study of the $\Lambda\Lambda$ and $\Xi$ , {i.e.} $S=-2$ systems. In these nuclei, experimental information is extremely scarce
and we have to often rely on the naive quark model to extract the coupling 
constants. 

The paper is organized as follows. In section 2, we present a brief discussion of the FSU Gold Lagrangian density and the  
method followed for the description of the hypernucleus. We also discuss 
the parameters involved in the effective $YN$ an $YY$ forces, and the procedure applied to determine them. Section 3 is 
dedicated to our results. Finally, in section 4, we summarize our
conclusions.

\section{Formalism}
\label{sec:1}
RMF calculations have been able to explain different features 
of stable and exotic nuclei like ground state binding energy, deformation,
radius, exited states, spin-orbit splitting, neutron halo etc.\cite{16} 
There are a number of different Lagrangian densities as well as a number of 
different parametrization. In the present work the FSU Gold Lagrangian density
 has been employed.\cite{17}
 While similar in spirit to most other forces, it contains 
two additional non-linear meson-meson interaction terms in the Lagrangian 
density, whose main virtue is a softening of both the EOS of symmetric matter 
and the symmetry energy. 
As a result, the new parametrization becomes more 
effective in reproducing a few nuclear collective modes,\cite{17} namely the breathing 
mode in $^{99}$Zr and $^{208}$Pb, and the isovector giant dipole resonance in
$^{208}$Pb. However, to the best of our knowledge, the effectiveness of this force in the 
hypernucleonic sector has not yet been established. Motivated by this fact we 
study single and double $\Lambda$ hypernuclei, as well as $\Xi$ hypernuclei using FSU Gold force.
We compare some of our results
 with those obtained with the NLSH force.\cite{18} 

\subsection{Model}
The standard FSU Gold Lagrangian density is given by the following form:

\begin{eqnarray}
\mathcal{L}_{total} &=& \bar{\psi}(i\gamma_{\mu}\partial^{\mu}-M_{n}){\psi}+\frac{1}{2}(\partial_{\mu}\sigma\partial^{\mu}\sigma-m_{\sigma}^{2}\sigma^{2}) \nonumber 
-\frac{1}{4}\Omega_{\mu\nu}\Omega^{\mu\nu}+\frac{1}{2}m_{\omega}^{2}\omega_{\mu}\omega^{\mu}\\
&&-\frac{1}{4}\vec{\rho}_{\mu\nu}\cdot\vec{\rho}^{\mu\nu} 
+\frac{1}{2}m_{\rho}^{2}\vec{\rho}_{\mu}\cdot\vec{\rho}^{\mu}-\frac{1}{4}A_{\mu\nu}A^{\mu\nu}\\&&
+g_{\sigma n}\bar{\psi}\psi\sigma \nonumber 
-\bar{\psi}\gamma_{\mu}(g_{\omega n}\omega^{\mu}+\frac{g_{\rho n}}{2}\vec{\tau}\cdot\vec{\rho}^{\mu}+\frac{e}{2}A_{\mu}(1+\tau_{3}))\psi \nonumber \\
&&-\frac{k}{3!}(g_{\sigma n}\sigma)^{3}-\frac{\lambda}{4!}(g_{\sigma n}\sigma)^{4}+\frac{\zeta}{4!}(g_{\omega n}^{2}\omega_{\mu}\omega^{\mu})^{2} \nonumber 
+\Lambda_{v}(g_{\rho n}^{2}\vec{\rho}_{\mu}\cdot
\vec{\rho}^{\mu})(g_{\omega n}^{2}\omega_{\mu}\omega^{\mu})
\end{eqnarray}
where
\begin{eqnarray}
  A_{\mu\nu} &=& \partial_{\mu}A_{\nu}-\partial_{\nu}A_{\mu}\nn\\
\Omega_{\mu\nu}&=&\partial_{\mu}\omega_{\nu}-\partial_{\nu}\omega_{\mu}\\
\vec{\rho}_{\mu\nu} &=&\partial_{\mu}\vec{\rho}_{\nu}-\partial_{\nu}\vec{\rho}_{\mu}-g_{\rho}(\vec{\rho}_{\mu}X\vec{\rho}_{\nu})\nn
\end{eqnarray}
The different terms are the standard ones used by Todd-Rutel \etal\cite{17}. 
The suffix `{\it n}' in the coupling constants refers to nucleons.

The hyperon couple to the non-strange mesons as well as the strange mesons
$\sigma^\ast$ and $\phi$.
Hence, the hyperon is introduced into the system through the hypernuclear sector of the Lagrangian density 
by adding the following term to the Lagrangian density in~(1). 
\begin{eqnarray}
\mathcal{L}_{h} &=& \bar{\psi}_{h}[i\gamma_{\mu}\partial^{\mu}-M_{h}+g_{\sigma h}\sigma+ 
g_{\sigma^{\ast}h}\sigma^{\ast}\nn\\&&-g_{\omega h}\gamma^{\mu}\omega_{\mu}-g_{\rho h}
\gamma^{\mu}\vec{\tau}\cdot\vec{\rho}^{\mu}
-g_{\phi h}\gamma^{\mu}\phi_{\mu}+\frac{e}{2}\gamma^\mu A_\mu(1+\tau_3)]{\psi}_{h} 
\end{eqnarray}
Here $\psi_{h}$ represents a hyperon field and $M_h$ is the mass of the hyperon. The 
coupling to the mesons are given by the coupling constants with a suffix `$h$'. The $\Lambda$ hyperon being charge-neutral and isoscalar does not couple to the 
photon or the $\rho$ meson.  While solving the coupled equations, the hyperon
contributions to the source terms for all the meson fields are taken into 
consideration, thus allowing for rearrangement of the normal nucleonic core. 






Pairing is introduced under the BCS approximation using a non zero range 
pairing force of strength $300$ MeV-fm for both protons and neutrons. The 
RMF-BCS equations are solved under the usual assumptions of classical mean 
fields, time reversal symmetry, no-sea contribution etc. Solutions of the Dirac
 and Klein-Gordon equations has been obtained directly in coordinate space.
Spherical symmetry is assumed for all the nuclei considered here.

\subsection{Meson Parameters for the $\Lambda N$ and $\Lambda\Lambda$ 
interactions}

In the naive quark model, it is assumed that the omega and the rho fields couple 
only to the $u$ and $d$ quarks, and the strange quark in the baryon acts  
as a spectator when coupling to the vector mesons. The vector meson-hyperon 
coupling constant ($g_{\omega h}$), is related to the vector meson-nucleon  
coupling constant $g_{\omega n}$ as 
\begin{eqnarray}\frac{1}{3}g_{\omega n} = \frac{1}{2}g_{\omega\Lambda} 
\end{eqnarray}
for a $\Lambda$ hypernuclei.\cite{19}
The scalar meson-hyperon coupling constant has often been determined by the
requirement to reproduce the potential depth of the hyperon in normal nuclear 
matter according to
 \begin{equation}
 U_h^n = g_{\sigma h}{\sigma}+g_{\omega h}\omega_0
\end{equation}
The potential depth for a $\Lambda$ in nuclear matter
has a well known value $U_{\Lambda}^{n} = -30$ MeV,\cite{20} which
 can be used to obtain $g_{\sigma\Lambda}$.

However, the usual procedure to extract the meson-nucleon coupling constants
and the meson masses in RMF approach involves reproducing not only the saturation properties of 
nuclear matter but also various properties of finite nuclei. Hence, the
nuclear matter properties may not be exactly reproduced. For example, 
the FSU Gold force predicts a binding energy per nucleon of $16.30$ 
MeV for symmetric nuclear matter while the commonly accepted value is
$16.0$ MeV. 

With the above idea in mind, the hyperon-meson coupling constants have been
fitted in the present work to reproduce the experimental hyperon binding energy
in case of single-$\Lambda$ hypernuclei.
One needs to remember that the mean field approximation may not work very well
in very light nuclei. Thus, only hypernuclei with $A\ge$16 were chosen for the 
fitting procedure. 
The contribution of $\sigma^{\ast}$ and $\phi$ mesons are taken 
into consideration for both single-$\Lambda$ and double-$\Lambda$
systems; these parameters are discussed later in this section.

We assume that the non-strange sector of the Lagrangian is 
completely determined and the corresponding parameters are adopted from the
work by Todd-Rutel \etal\cite{17} and
the only task that remains is fixing the hyperon-meson coupling constants. The 
mass of the $\Lambda$ has been fixed at 1115.6 MeV.
The initial values for the meson-hyperon coupling constants were taken from the 
naive quark model as described above. The two parameters were then
varied in order to fit the experimental hyperon separation energies.
Thus a best fit procedure was adopted to obtain the values of the two
coupling constants. 
However, it was observed that the vector meson-hyperon coupling constant
determined from the naive quark model is sufficient for fit. 
Consequently, only the value of the $g_\sigma\Lambda$ needed to be varied.
The best fit parameters and the fitted separation energies are presented 
in Tables 1 and 2, respectively.

It is, of course, imperative that the constants determined in the above procedure
 give quite a reasonable agreement with the nuclear matter properties. The present
values of the parameters give rise to a lambda potential depth of $-28.7$ MeV, 
close to the value of $-30.0$ MeV adopted by Schaffner \etal\cite{20}
 
Determining the $\Lambda\Lambda$ contribution to the experimental binding 
energy $B_{\Lambda\Lambda}$ is subject to large
uncertainty because of the scarcity of data  available on double-$\Lambda$
nuclei. 
Nuclear emulsion experiments reported the observation
of three double-$\Lambda$ hypernuclei: $^{6}_{\Lambda\Lambda}$He, $^{10}_{\Lambda\Lambda}$Be and 
$^{13}_{\Lambda\Lambda}$B. From these events, an effective
 $\Lambda\Lambda$ matrix element $-V_{\Lambda\Lambda} = \Delta B_{\Lambda\Lambda} = |B_{\Lambda\Lambda}| 
- 2|B_{\Lambda}| \cong 4-5$ MeV was determined,\cite{21} $|B_{\Lambda\Lambda}|$ 
being the separation energy of the $\Lambda$ pair from the 
$^{A}_{\Lambda\Lambda}$Z hypernucleus, given by

 
  \begin{eqnarray}  
   B_{\Lambda\Lambda}(^A_{\Lambda\Lambda}Z) &=& B(^A_{\Lambda\Lambda}Z)-B(^{A-2}Z) 
\end{eqnarray}
and $|B_{\Lambda}|$, the hyperon separation energy from the $^{A}_{\Lambda}$Z hypernucleus.


On the other hand, a very recent counter-emulsion hybrid experiment, performed at KEK,\cite{22} favours a quite weaker $\Lambda\Lambda$ interaction: 
$ \Delta B_{\Lambda\Lambda}(^{6}_{\Lambda\Lambda}He) = 0.67\pm 0.17$ MeV
from a recent reanalysis of the data\cite{50}. 

We take the meson masses equal to  
$m_{\sigma\ast} = 980$ MeV and $m_{\phi} = 1020$ MeV. For the
$\phi$ coupling we take the naive quark model value\cite{3} obtained from the relation
 \begin{equation} \frac{1}{3} g_{\omega n} = -\frac{1}{\sqrt 2}g_{\phi \Lambda} \end{equation}
while the $\sigma^{\ast}$ coupling strength is 
determined by reproducing the value of $\Delta B_{\Lambda\Lambda}$ for 
the nucleus $^{6}_{\Lambda \Lambda}He$ within error.

\subsection{Parameters for the $\Xi N$ interaction}
\label{sec:2}
In the naive-quark model, the relations between the vector meson-nucleon 
and the vector meson-hyperon coupling constants are given as,
\begin{eqnarray}\frac{1}{3}g_{\omega n} = g_{\omega\Xi^-} = g_{\omega\Xi^0} \end{eqnarray}
\begin{equation}
g_{\rho \Xi^-}=g_{\rho \Xi^0}=g_{\rho n}
\end{equation}
for a $\Xi$ hypernucleus.\cite{19}
Dover and Gal\cite{24} analyzed old emulsion data 
of $\Xi^{-}$ hypernuclei and obtained a nuclear potential well depth of $U_{\Xi} = -21$ to $-24$ MeV.
Fukuda \etal\cite{25} fitted the very low energy part of $\Xi^{-}$ hypernuclear spectrum in the 
$^{12}C(K^{-},K^{+})$X reaction and estimated the value of $U_{\Xi}$ to be between $-16$ to $-20$ MeV.
E885 at the AGS\cite{26} have indicated a potential depth of $U_{\Xi} = -14$ MeV or less.
Here, we choose $U_{\Xi^{-}} = U_{\Xi^{0}} = -16$ MeV initially to determine the parameters of effective $\Xi N$ couplings.
However, the corresponding parameters does not reproduce the empirical $\Xi^{-}$ separation energies. 
Therefore, we tune the parameters to match the experimental results slightly better 
(as there are not enough experimental data available to formally fit the data). The
parameters thus determined give a $\Xi$ potential depth of -20.59 MeV, which is closer
to the value -21.0 MeV determined from the old emulsion data.\cite{24}

The adopted values for all these parameters are presented in Table \ref{tab:1}. We have calculated separation 
energies of the single-$\Lambda$ systems using the NLSH force to make a comparison. The $\Lambda N$ coupling 
constants for the NLSH parameter set were fitted in the same procedure. 
Table 1 also lists the parameters used for calculations of single-$\Lambda$ separation energies with the force NLSH. 

\begin{table}[h]
\begin{center}
\caption{Model parameters used in this work. The hyperon masses are in MeV.
}
\label{tab:1}
\begin{tabular}{ccccccccccc}
\noalign{\smallskip}\hline
Model&
$M_\Lambda$ &
$g_{\sigma\Lambda}$& 
$g_{\omega\Lambda}$ &
$g_{\sigma^{\ast}\Lambda}$&
$g_{\phi\Lambda}$ &   
$M_\Xi$ & 
$g_{\sigma\Xi}$       &      
$g_{\omega\Xi}$       &       
$g_{\rho\Xi}$         \\\hline     
FSU Gold &1115.6 & 6.519 & 9.530 & 6.515 & -6.740&
1670 &3.471 & 4.767 & 5.884 \\
NLSH & 1115.6 & 6.465 &8.630 & - & - & - & - & -\\
\noalign{\smallskip}\hline
\end{tabular}
\end{center}
\end{table}
\section{Results}
\subsection{Single $\Lambda$ hypernuclei: ground states} 
\begin{table}[h]
\begin{center}
\caption{
Binding energy per nucleon ($BE/A$) and separation energy ($S_Y$) of $\Lambda$ 
hyperon calculated for the hypernuclei included in the fitting 
procedure.  All energy values are in MeV.
Experimental values, unless otherwise indicated, are from 
Hashimoto \etal\cite{1}}
\begin{tabular}{cccccl}\hline
Nuclei               &\multicolumn{2}{c}{FSU Gold}    &\multicolumn{2}{c}{NLSH} &\multicolumn{1}{c} {Exp.}\\\cline{2-6}
                     &$BE/A$    & $S_{Y}$ &$BE/A$   & $S_{Y}$ & $~~~~S_{Y}$   \\\hline
$^{16}_{\Lambda}$O   & 7.79    & 12.21   &7.95    &12.29  & 12.42(05)\cite{27}\\
$^{17}_{\Lambda}$O   & 8.23    & 12.51   &8.33    &12.52  & 13.39(55)\\
$^{28}_{\Lambda}$Si  & 8.19    & 17.49   &8.41    &18.05  & 16.60(20)\\
$^{32}_{\Lambda}$S   & 8.29    & 18.26   &8.44    &18.55  & 17.50(50)\cite{31}\\
$^{33}_{\Lambda}$S   & 8.42    & 18.55   &8.53    &18.59 & 17.96(000)\cite{32}\\
$^{40}_{\Lambda}$Ca  & 8.62    & 18.45   &8.64    &18.59 & 18.70(110)\cite{33}\\
$^{41}_{\Lambda}$Ca  & 8.79    & 18.57   &8.79    &18.68 & 19.24(100)\\
$^{51}_{\Lambda}$V   & 8.85    & 20.37   &8.96    &20.90  & 19.97(100)\cite{35}\\
$^{56}_{\Lambda}$Fe  & 8.83    & 21.18   &8.97    &21.69  & 21.00(100)\\
$^{89}_{\Lambda}$Y   & 8.84    & 22.62   &8.89    &23.19  & 23.10(50)\\ 
$^{139}_{\Lambda}$La & 8.54    & 24.59   &8.60    &24.68  & 24.50(120)\\  
$^{208}_{\Lambda}$Pb & 7.98    & 25.12   &8.03    &26.02  & 26.30(86)\\\hline
\end{tabular}
\end{center}
\end{table}

In Table 2, the results of our calculation for the single-$\Lambda$ hypernuclei,
included in the fitting procedure, are presented. We tabulate the total energy
per nucleon as well as the separation energy ($S_{Y}$), the latter being 
compared with experimental values. The results for the NLSH Lagrangian density are also tabulated.
 The hyperon is placed in the $1s_{1/2}$ state in all the cases.
One can see that the values come very close to the experimental measurements. 
Except in the case of  $^{208}_{\Lambda}$Pb, the values differ by less than 1 MeV. 
 It is also easy to see that 
FSU Gold gives a better agreement than NLSH, since the rms deviation for FSU Gold comes out to be 
0.64 MeV, whereas the same for NLSH is 0.72 MeV. 

With the success of this model in $A\ge16$, we have also extended our calculations for 
lighter hypernuclei. The results are presented in Table \ref{tab:3}. We find 
that though the
errors are slightly larger, the present approach can reproduce the hyperon separation energy
to a reasonable degree. 
\begin{table}[h]
\begin{center}
\caption{
Binding energy per nucleon ($BE/A$) and separation energy of $\Lambda$ hyperon ($S_{Y}$) for
hypernuclei with $A < 16$. All energy values are in MeV. Hyperons are placed in $1s_{1/2}$ state.
 Experimental values are taken from Bando \etal\cite{39}.
\label{tab:3}       
}
\begin{tabular}{lcll}\hline
Nuclei      &\multicolumn{2}{c}{Present work} &\multicolumn{1}{c}{Exp.}\\\cline{2-4}
                     &$BE/A$   & $~~S_{Y}$  & $~~~~S_{Y}$\\\hline
$^{8}_{\Lambda} $Be  & 5.38   & ~5.53       & ~6.84(05)  \\
$^{10}_{\Lambda}$Be  & 6.16   & ~8.11      & ~9.11(22)  \\
$^{11}_{\Lambda}$B   & 6.55   & ~9.29       & 10.24(05)  \\
$^{12}_{\Lambda}$B   & 7.04   & 10.49    & 11.37(06) \\
$^{12}_{\Lambda}$C   & 6.79   & 10.44     & 10.76(19)  \\
$^{14}_{\Lambda}$C   & 7.62   & 11.78      & 12.17(33) \\
$^{14}_{\Lambda}$N   & 7.39   & 11.76     & 12.17(000) \\
$^{15}_{\Lambda}$N   & 7.74   & 11.95    & 13.59(15)  \\\hline
  
\end{tabular}
\end{center}
\end{table}

In Fig. 1 we present the results of our calculations for the $\Lambda$ 
separation energies in various shells for a number of 
hypernuclei with mass scale and compare them with the experimental values.
The relevant graph for the present discussion corresponds to the $s_\Lambda$ shell.
\begin{figure}[h]
\center
\resizebox{10cm}{!}{
\includegraphics{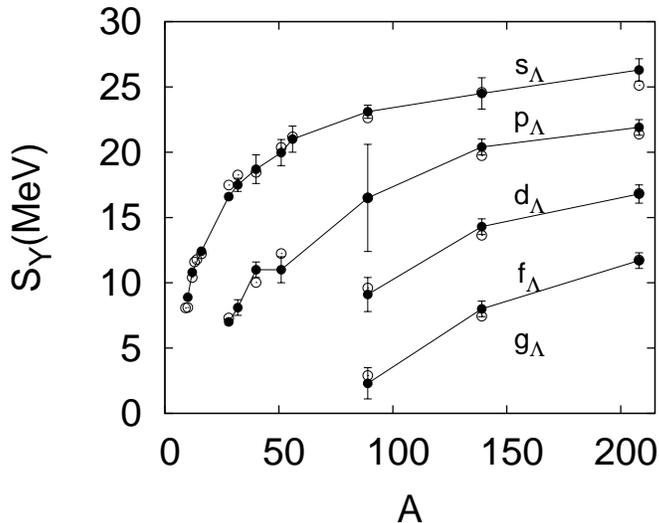}
}
\caption{
The separation energies of $\Lambda$ in different single-$\Lambda$ hypernuclei as a function of the 
baryon number A. Filled (empty) circles indicate experimental (theoretical) 
values. The lines are only for ease of visualization.
\label{fig:1}}
\end{figure}
We note that the maximum deviation from experimental data occurs
in case of $^{208}_{\Lambda}$Pb. However, in this case also, the under-estimation is smaller in 
our calculations than SHF calculations.\cite{14,37} 
RMF calculations by Mi-Xiang \etal\cite{38} seems to underestimate the binding 
energy to varying magnitudes in all heavier systems, leading Guleria
\etal\cite{14} to conclude that RMF under-predicts 
the binding energy in hypernuclei 
above $A= 87$ as compared to SHF calculations. Our calculation, however,
strongly disagrees with this conclusion, as the agreement between 
experimental data and the results of this calculation is consistent throughout 
the periodic table. Compared to our calculation, RMF calculation\cite{5} using 
NLSH parameters either under-predicts or over-predicts the binding energy 
except for the case of $^{17}_{\Lambda}$O.
 However, even with the few values reported there, it is possible 
to note that the calculations we report are in much better agreement with 
experiment\cite{5}.

\begin{table}[h]
\label{tab:4}
\begin{center}
\caption{
Binding energy per nucleon ($BE/A$) (in MeV)  predicted for a number of
hypernuclei. Hyperons are placed in $1s_{1/2}$ state. All energy values are in MeV. See text for details.
 } 
\begin{tabular}{clll|clll}
 \hline
$_\Lambda^AZ$     &     \multicolumn{3}{c|}{BE/A}  &     $_\Lambda^AZ$      &     \multicolumn{3}{c}{BE/A}\\\cline{2-4}\cline{6-8}
& Pres.  & RMF\cite{38}  & SHF\cite{14} &   &   Pres. & RMF\cite{38} & SHF\cite{14} \\\hline
$^{15}_{\Lambda}$N   & 7.746 &-        & 8.623  &$^{88}_{\Lambda}$Sr&   8.849    &7.657    & 8.762\\
$^{20}_{\Lambda}$Ne  & 7.772 &7.635    & 7.445  &$^{89}_{\Lambda}$Sr&   8.862    &7.564    & 8.742 \\
$^{24}_{\Lambda}$Mg  & 7.832 &7.723    & 7.671  &$^{90}_{\Lambda}$Y &   8.854    &7.510    & 8.746  \\
$^{27}_{\Lambda}$Al  & 8.204 &-        & 8.225  &$^{112}_{\Lambda}$Sn&   8.628    &6.925    & 8.477 \\
$^{33}_{\Lambda}$S   & 8.422 &-        & 8.706  &$^{117}_{\Lambda}$Sn&   8.624    &7.028    & 8.515\\
$^{36}_{\Lambda}$S   & 8.666 &8.637    & 8.613  &$^{120}_{\Lambda}$Sn&   8.599    &7.026    & 8.535 \\                                              
$^{49}_{\Lambda}$Ca  & 8.827  &8.880    & 8.767  &$^{136}_{\Lambda}$Xe & 8.492   &6.949    & 8.397   \\
$^{55}_{\Lambda}$Fe  & 8.840 &8.797    & 8.731  & $^{140}_{\Lambda}$La & 8.403   &6.893    & 8.366   \\
$^{60}_{\Lambda}$Ni  & 8.782 &8.841    & 8.751  &$^{143}_{\Lambda}$Pm&   8.390    &-        & -    \\
$^{86}_{\Lambda}$Kr& 8.844   &8.778    & 8.735  &$^{145}_{\Lambda}$Sm&   8.360    &6.790    & 8.294\\
$^{87}_{\Lambda}$Kr&8.848    &8.796    & 8.702  &$^{209}_{\Lambda}$Pb&   7.982   &6.726  & 7.888 \\
$^{88}_{\Lambda}$Rb&8.859    &7.587    & 8.727  &$^{210}_{\Lambda}$Bi&   7.970     &6.711    & 8.857 \\

\hline
\end{tabular}
\end{center}
\end{table}

In Table 4 we present the results of our calculations (Pres.) of 
binding energy per nucleon for a 
number of hypernuclei over a wide range of mass number. We see that for hypernuclei with baryon number in the range
of 35-95 the BE/A is around 8.7 MeV and it decreases gradually with further increase in the mass
number. The present calculation agrees with the SHF calculations \cite{14} (SHF)
to a great extent while the RMF calculations by Mi-Xiang \etal\cite{38} (RMF) seems to fail beyond $A=87$, as already noted.
             
\subsection{Single $\Lambda$ hypernuclei: excited states}

\begin{figure}[h]
\center
\resizebox{8cm}{!}{ 
\includegraphics{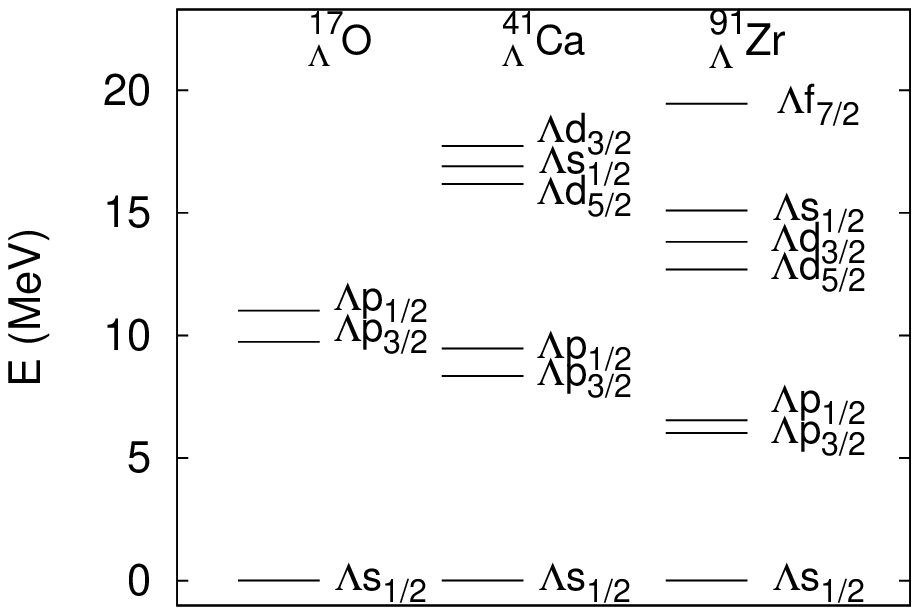}}
\caption{
Excitation energy in some odd mass $\Lambda$ hypernuclei.}
\label{fig:2}  
\end{figure}

We also study the excited states that can be occupied by the $\Lambda$
in hypernuclei. For this, we calculate self-consistent solutions for 
the ground state and the excited states separately and subtract to obtain 
the excitation energies. Our first interest lies in the nuclei which have closed 
nucleon core and one $\Lambda$-hyperon. The results for excitation energy in $^{17}_{\Lambda}$O, $^{41}_{\Lambda}$Ca
and $^{91}_{\Lambda}$Zr with respect to the hypernuclear ground state are presented in Fig. \ref{fig:2}. 

Unfortunately, no results are available in any of the above nuclei.
In fact, most of the nuclei, where the energy values corresponding to excited 
states of $\Lambda$ hyperon are known, are of even mass number. 
In such nuclei, the  $\Lambda$ hyperon couples to the odd nucleon 
(neutron or proton). The residual interaction between the hyperon and the
nucleon is not considered at the mean field level. However, the energy 
separation between states arising out of the above coupling
is much smaller compared  to the excitation energy. In fact, due to the 
limitations of experimental resolution, in most cases the individual
states have not at all been observed. 
In Table 5, we summarize the
available experimental information on excitation energy and compare them with our calculation. One can see that
the theory describes the excitation energies reasonably well, considering the 
fact that the residual interaction has been completely ignored. We also present
a number of nuclear states where experimental information is not yet available.
Fig. 1 also presents the hyperon separation energies for the excited states.

\begin{table}
\begin{center} thee
\label{tab:5}
\caption{\label{exc}Excitation energy (in MeV) of different
$\Lambda$ states. Energy values marked with `*' are given for $l$ 
excitations in literature. The superscript to the right of the hypernucleus
in column 1 indicates the reference from which the experimental values have been obtained. 
}
\begin{tabular}{llrrlrrccccc}
\hline
Nucleus & $\Lambda$-state & \multicolumn{2}{c}{Excitation Energy}
 & $\Lambda$-state & \multicolumn{2}{c}{Excitation Energy}\\\hline
&&\multicolumn{1}{c}{Exp.} & \multicolumn{1}{r}{Theo.}&&\multicolumn{1}{c}{Exp.} & \multicolumn{1}{r}{Theo.} \\\hline
$^{12}_\Lambda$B\cite{40} & $p_{3/2}$ & 10.93 &10.40 &\\
$^{12}_\Lambda$C\cite{35} & $p_{3/2}$ &10.66 & 10.34   &\\
$^{13}_\Lambda$C\cite{41} & $p_{3/2}$ & 9.93 &11.02 &\\
$^{16}_\Lambda$O\cite{42} & $p_{3/2}$ & 9.1~ &9.97 
                 & $p_{1/2}$ & 11.0 & 11.21\\
$^{28}_\Lambda$Si\cite{43} &$p_{3/2}$ & 9.6~ & 10.18  &\\
$^{51}_\Lambda$V\cite{35} & $p_{3/2}$ & 8.07& 7.92    
        & $p_{1/2}$ & 9.40& 8.73   &\\
        & $d_{5/2}$ &16.42& 15.93   
        & $d_{3/2}$ &18.42& 17.51  &\\
$^{89}_\Lambda$Y\cite{35}  &$p_{3/2}$ & 6.01& 6.10 
        & $p_{1/2}$ & 7.38& 6.59   &\\
        & $d_{5/2}$ &12.79& 12.81   
        & $d_{3/2}$ &14.42& 13.93   \\
        & $f_{7/2}$ &19.98& 19.56   
        & $f_{5/2}$ &21.68& 21.19  \\
$^{139}_\Lambda$La\cite{1} &$p_{3/2}$ & 4.1$^*$ &4.86 &$p_{1/2}$ & &5.64\\
                  &  $d_{5/2}$ & 10.2$^*$& 10.95 
                  & $d_{3/2}$ & & 11.01\\
                  & $f_{7/2}$ &16.5$^*$ & 17.15
                  & $f_{5/2}$ & & 19.38\\
$^{208}_\Lambda$Pb\cite{43} &$p_{3/2}$ & 5.2~&3.74
                   & $p_{1/2}$ & & 3.91 \\
                   &  $d_{5/2}$ &  &8.28 
                   &  $d_{3/2}$ &  &8.71\\
                   & $f_{7/2}$ &  &13.33
                   & $f_{5/2}$ &  &14.19\\
\hline
\end{tabular}
\end{center}
\end{table}

It should be noted that the origin of the states in $^{16}$O has not been 
discussed by Agnello \etal\cite{42}. However, it is clear that the lowest 
observed excited state at 6.1 MeV does not correspond to the excited state of 
$\Lambda$. This has also been supported by previous measurements.
We should also point out that Hasegawa \etal\cite{43} have identified the 
$\Lambda$-excitation energy with the major shell $p$ only. However, 
comparison with later experiments and  with theoretical results derived in the 
present work, allow us to identify it with the $p_{3/2}$ state. For example, 
in $^{89}_\Lambda$Y, the energy of the  major shell $p$ has been measured in
Hasegawa \etal\cite{43} as 5.9(6) MeV while Hotch \etal\cite{35} have 
measured the energies of $p_{3/2}$ and $p_{1/2}$ as 6.01 and 7.38 MeV 
respectively. In some other cases also, experimental data are available for 
$l$ excitations. In such situations we have always compared them with the 
lower of the two $j$ states originating from the angular momentum state.

One should remember that the various $J$ levels arising out of the coupling of 
the neutron hole to the $\Lambda$ are degenerate in the mean field level.
If the single particle state occupied by the  ordinary nucleon is 
experimentally known, we have used the tagging method to put the last nucleon
in that state in our calculation. Also notable is the small spin-orbit coupling for the $\Lambda$ states. 
Although our calculations slightly underestimates
this difference, the trend is generally reproduced in agreement with the experimental data (see Table 5) as well as 
the previous theoretical works.\cite{2a,2b}

\subsection{Double $\Lambda$ hypernuclei}

We calculate binding energies for several double-$\Lambda$ hypernuclei including light, medium, and heavy
systems within our framework using the parameter set FSU Gold.
Our results (Pres.) are presented in Table 6 for the $\Lambda\Lambda$ binding energy $B_{\Lambda\Lambda}$ 
and the quantity $\Delta B_{\Lambda\Lambda}$. 
The experimental data are also listed, where available, for comparison. 
Results from another RMF calculation\cite{3} using NLSH are also presented.

 



\begin{table}[h]
\label{tab:6}
\begin{center}
\caption{
$B_{\Lambda\Lambda}$ and $\Delta B_{\Lambda\Lambda}$ of double-$\Lambda$ hypernuclei. The available 
experimental data\cite{22,44,45,46,47} are also presented. All energy values are in MeV.
See text for details.}
\begin{tabular}{lccc|ccc}
\hline
Nuclei       &\multicolumn{3}{c|}{$B_{\Lambda\Lambda}$} &\multicolumn{3}{c}{$\Delta B_{\Lambda\Lambda}$}\\\cline{2-7}
  &                        Exp.       &Pres.  &NLSH   &Exp.    &Pres.  &NLSH\\
 \hline
$^{6}_{\Lambda\Lambda}$He  &7.25$\pm$0.2 &  ~3.98  &~4.68          &0.67$\pm$0.17  &      0.62           &1.01   \\      
$^{10}_{\Lambda\Lambda}$Be &17.7$\pm$0.7&   14.03 &15.94         &4.3$\pm$0.4  &      0.27           &0.29  \\ 
$^{13}_{\Lambda\Lambda}$B  &27.5$\pm$0.7&21.17        &22.52         &4.8$\pm$0.7  & 0.19        &0.21  \\
$^{18}_{\Lambda\Lambda}$O  &          &26.96         &25.12         &           & 0.11         &0.07   \\        
$^{42}_{\Lambda\Lambda}$Ca &          &37.75         &37.17         &           &0.03        &0.00   \\     
$^{210}_{\Lambda\Lambda}$Pb&          &50.34         &53.02         &           & 0.02        &0.02    \\
\hline
   
\end{tabular}
\end{center}
\end{table}
It is seen that 
$B_{\Lambda\Lambda}$ increases whereas $\Delta B_{\Lambda\Lambda}$ decreases with increase 
in mass number. This observation is in agreement with previous works.\cite{4,49}
Our result for $^{6}_{\Lambda\Lambda}$He matches with the NAGARA event\cite{22} result from a recent 
reanalysis of the data\cite{50}  within experimental error. 
The results for $\Delta B_{\Lambda\Lambda}$
of $^{10}_{\Lambda\Lambda}$Be
and $^{13}_{\Lambda\Lambda}$B do not, however, match with the empirical data. This is reasonable as 
the empirical data signifies a strong $\Lambda\Lambda$ interaction, which is now-a-days believed 
to be wrong in view of the NAGARA event results. The $\Delta B_{\Lambda\Lambda}$ for all the nuclei 
presented here agrees well with the results of Shen \etal\cite{3}. 
However, we should point out that the quantity $\Delta B_{\Lambda\Lambda}$
decreases very rapidly with increase in mass. Thus, in heavier nuclei, this 
quantity becomes so small that the error related to the convergence of the 
mean field solutions may become comparable to it.


As pointed out by Marcos \etal\cite{4} RMF theory cannot compete
with more elaborate three-body calculations for $\Delta B_{\Lambda\Lambda}$.
 In particular for such a light system as $^{6}_{\Lambda\Lambda}$He, its
application is questionable. In view of extensions to multi-$\Lambda$
systems, however, it is important to check the constraints it
brings on the coupling of the $\Lambda$ to the various meson fields.
 In this respect, it would be very desirable to obtain more and better
experimental data for heavier hypernuclei.

The influence of $\Lambda$ hyperons on the nuclear core,
which is known as the core polarization effect\cite{51}, is an interesting 
aspect of hypernuclei as the nucleons in the core are affected by the 
additional $\Lambda$ hyperons in hypernuclei. 
The so-called rearrangement energy quantifies the core polarization
effect, which represents the change of nuclear core binding energies caused by the
presence of $\Lambda$.
We present in Table 7 the rearrangement energy in several single and 
double-$\Lambda$ hypernuclei.


\begin{table}[h]
\label{tab:7}
\begin{center}
\caption{
Rearrangement energies in MeV ($E_R$) in several single and double-$\Lambda$ 
hypernuclei. We also present the absolute value of the single particle energy 
of the $\Lambda$ in the $1s_{1/2}$ state.
}
\begin{tabular}{lcl}
\hline
Nucleus&  $E_{R}$  &  $\epsilon_{\Lambda}$(1s)\\\hline
$^{17}_{\Lambda}$O      &  0.320 &  12.827  \\
$^{18}_{\Lambda\Lambda}$O      & 0.701   & 12.912 \\\hline
$^{41}_{\Lambda}$Ca  &  0.127  &  18.699  \\
$^{42}_{\Lambda\Lambda}$Ca  &  0.306  & 18.738 \\\hline
$^{209}_{\Lambda}$Pb    & 0.048   &  25.211  \\
$^{210}_{\Lambda\Lambda}$Pb   &  0.100   & 25.222 \\\hline
\end{tabular}
\end{center}
\end{table}
It is seen that the rearrangement energy decreases
rapidly with increasing mass number, and is usually negligible in comparison 
with the binding energy except for very light systems. The rearrangement energy
does not appear to be a linear function of the number of hyperons.

\subsection{Single $\Xi$ hypernuclei}
\begin{figure}[h]
\center
\resizebox{10cm}{!}{ 
\includegraphics{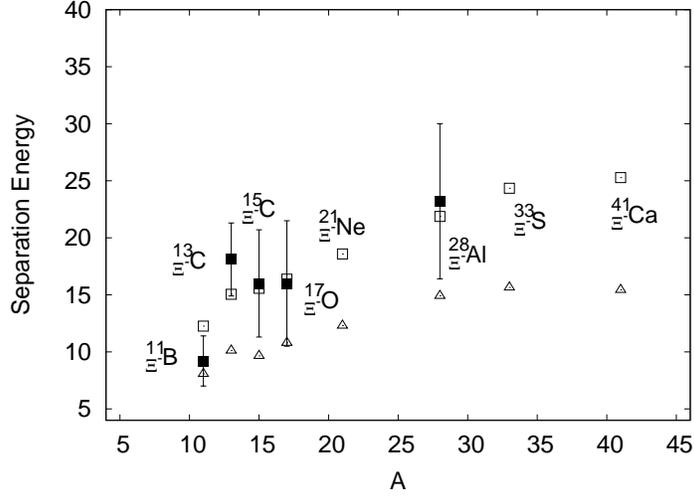}
}
\caption{
Separation energy(in MeV) of $\Xi$ hyperon in hypernuclei. Filled (empty) boxes represent experimental
 (theoretical) $\Xi^{-}$ separation energies while
empty triangles represent theoretical $\Xi^{0}$ separation energies. }
\label{fig:3}  
\end{figure}

The results of calculations for the $\Xi$ hypernuclear systems are presented in the Fig. 3.
 We find that, except  in one case, the energy value has
been predicted accurately  within experimental errors. The trend is also
generally reproduced. Of course, the experimental errors are rather large and
better values may be obtained when more accurate measurements are available. 
However, it is clear from the figure that our calculations reproduce the experimental results quite satisfactorily, including the kink
at $^{13}_\Xi$C.

\section{Summary and conclusion}

To summarize, the FSU Gold Lagrangian density has been extended to include 
hypernuclei. The meson-hyperon coupling constants have been varied to
reproduce the hyperon binding energy whenever sufficient experimental data are
available. Otherwise, the naive quark model has been invoked to fix the 
parameters. The new parameters can reproduce the potential depth of the hyperons
in nuclear matter. The hyperon separation energies are also reasonably 
predicted, even in very light hypernuclei. Ground states in single and doubly 
strange hypernuclei for a wide range of mass number and excited states in singly strange hypernuclei have been studied in the new extended Lagrangian density 
with reasonable success. 
The present calculation works well in all the observed 
hypernuclear systems. New and improved experimental data are  
desirable in order to further verify the predictions of this model. 
With the success of this model for the $S= -1$ and 
$-2$ systems, we would like to extend it to the study of multi-strange
 exotic systems. This work is in progress.

\section*{Acknowledgements}

This work was carried out with financial assistance of the UGC (RFSMS, DRS, 
UPE). We sincerely thank the referee for the extremely useful comments and 
suggestions in improving the manuscript.


\begin{thebibliography}{99}
\bibitem{1} O. Hashimoto and H. Tamura, {\it Prog. Part. Nucl. Phys.} {\bf 57} (2006) 564.
 \bibitem{2a} W. Bruckner \etal, {\it Phys. Lett.} {\bf 79B} (1978) 157.
\bibitem{2b} S. Ajimura \etal, {\it Phys. Rev. Lett.} {\bf 86} (2001) 4255 and references therein.
\bibitem{3} H. Shen, F. Yang and H. Toki, {\it arXiv: [nucl-th]} 0602046v1 (2006).
\bibitem{4} S. Marcos, R.J. Lombard and J. Mares, {\it Phys. Rev. C} {\bf 57} (1998) 1178.
\bibitem{5} Y.-H. Tan, X.-H. Zhong, C.-H. Cai and P.-Z. Ning, {\it \PR C} {\bf 70} (2004) 054306.
\bibitem{6} http://j-parc.jp/researcher/Hadron/en/pac\_1201/minutes-j-parc-pac-20120115-final.pdf 
\bibitem{7a} H. Muller, {\it Phys. Rev. C} {\bf 59} (1999) 1405. 
\bibitem{7b} P. Papazoglou \etal, {\it Phys. Rev. C} {\bf 57} (1998) 2576.
\bibitem{8a} K. Saito, K. Tsushima and A. Thomas, {\it Prog. Part. Nucl. Phys.} {\bf 58} (2007) 1.
 \bibitem{8b} K. Tsushima, K. Saito, J. Heidenbauer and A.W. Thomas, {\it Nucl. Phys. A} {\bf 630} (1998) 691.
\bibitem{8c} P.M. Guichon, A.W. Thomas and K. Tsushima, {\it Nucl. Phys.} {\bf A 814} (2008) 66.
\bibitem{9} C.M. Keil, F. Hoffmann and H. Lenske, {\it Phys. Rev. C} {\bf 61} (2000) 064309.
\bibitem{10} D.J. Millener, {\it Nucl. Phys.} {\bf A 835} (2010) 11c and references therein.
\bibitem{11a} S. Iwao, {\it Prog. Theo. Phys.} {\bf 46} (1971) 1407.
\bibitem{11b} M. Grypeos, G. Lalazissis and S. Massen, {\it Nucl. Phys.} {\bf A 450} (1986) 283c.
\bibitem{11c} C. Samanta, P. Roy Chowdhury and D. N. Basu, {\it J. Phys. G:Nucl. Part. Phys.} {\bf 32} (2006) 363.
\bibitem{12a} R. Hausmann and W. Weise, {\it Nucl. Phys.} {\bf A 491} (1989) 598.
\bibitem{12b} D.J. Millener, C. B. Dover and A. Gal, {\it Phys. Rev. C} {\bf 38} (1988) 2700.
\bibitem{13} M. Rayet, {\it Nucl. Phys.} {\bf A 367}, (1981) 381 and references therein.
\bibitem{14} N. Guleria, S. K. Dhiman and R. Shyam, {\it arXiv:[nucl-th]} {\bf 1108.0787v1} 2011.
\bibitem{15a} M. Rufa \etal, {\it Phys. Rev. C} {\bf 42}
(1990) 2469.
\bibitem{15b} N. K. Glendenning \etal, {\it Phys. Rev. C} {\bf 48} (1993) 889.
\bibitem{15c} D. Vretenar, W. Poschl, G.A. Lalazissis and P. Ring, {\it Phys. Rev. C} {\bf 57} (1998) 1060.
\bibitem{17} B.G. Todd-Rutel and J. Piekarewicz, {\it \PRL} {\bf 95} (1995) 122501.
\bibitem{16}P. Ring, {\it Prog. Part. Nucl. Phys.} {\bf 37} (1996) 193.
\bibitem{18} M.M. Sharma, M.A. Nagarajan and P. Ring, {\it \PL} {\bf B312}, (1993) 377 .
\bibitem{19} K. Tsushima and F.C. Khanna, {\it Phys. Rev. C} {\bf 67} (2003) 015211,
{\it Prog. Theor. Phys. Suppl.} {\bf 149} (2003) 160.
\bibitem{20} J. Schaffner \etal,  {\it Ann. Phys. (N.Y.)} {\bf 235} (1994) 35.
\bibitem{21} B.F. Gibson, I.R. Afnan, J.A. Carlson and D. R. Lehman, {\it Progr. Theor.
Phys. Suppl.} {\bf 117} (1994) 339.
\bibitem{22} H. Takahashi \etal, {\it Phys. Rev. Lett.} {\bf 87} (2001) 212502.
\bibitem{50} K. Nakazawa, {\it Nucl. Phys.} {\bf A 835} (2010) 207.
\bibitem{24} C.B. Dover and A. Gal, {\it Ann. Phys.(N.Y)} {\bf 146} (1983) 309.
\bibitem{25} T. Fukuda \etal, {\it \PR C} {\bf 58} (1998) 1306.
\bibitem{26} P. Khaustov \etal, {\it \PR C} {\bf 61} (2000) 054603.
\bibitem{27} M. Ukai \etal, {\it Phys. Rev. Lett.} {\bf 93} (2004) 232501.
\bibitem{31} R. Bertini \etal, {\it Phys. Lett.} {\bf 83 B} (1979) 306.
\bibitem{32} G.A. Lalazissis, M.E. Grypeos and S.E. Massen, {\it \PR C} {\bf 37} (1988) 2098.
\bibitem{33} H. Tamura \etal, {\it Prog. Theo. Phys.} {\bf 117} (1994) 1.
\bibitem{35} H. Hotchi \etal, {\it Phys. Rev. C} {\bf 64} (2001) 044302.
\bibitem{37} J. Cugnon, A. Lejeune and H.J. Schulze, {\it Phys. Rev. C} {\bf 62} (2000) 064308.
\bibitem{38} L. Mi-Xiang, L. Lei  and N. Ping-Zhi, {\it Chin. Phys. Lett.} {\bf 26} (2009) 072101.
\bibitem{39} H. Bando, T. Motoba and J. Zotka , {\it Int. J. Mod. Phys. A} {\bf 5} (1990) 4021.
\bibitem{40} M. Iodice \etal, {\it \PRL} {\bf 99} (2007) 052501.
\bibitem{41} S. Ajimura \etal, {\it Nucl. Phys.} {\bf A 639} (1998) 93.
\bibitem{42} M. Agnello \etal, {\it \PL B} {\bf 698} (2011) 219.
\bibitem{43} T. Hasegawa \etal, {\it \PR C} {\bf 53} (1996) 1210.
\bibitem{44} M. Danysz \etal, {\it Nucl. Phys.} {\bf 49} (1963) 121.
\bibitem{45} D.J. Prowse \etal, {\it \PRL} {\bf 17} (1966) 782.
\bibitem{46} S. Aoki \etal, {\it Prog. Theor. Phys.} {\bf 85} (1991) 1287.
\bibitem{47} G.B. Franklin, {\it Nucl. Phys.} {\bf A 585} (1995) 83c.
\bibitem{49} D.E. Lanskoy, Y.A. Lurie and A.M. Shirokov, {\it Z. Phys. A} 
{\bf 357} (͑1997͒) 95.
\bibitem{51} J. Mares and J. Zofka, {\it Z. Phys. A} {\bf 333} (1989) 209.



\end{thebibliography}
\end{document}